\documentclass[runningheads]{llncs}
\usepackage[T1]{fontenc}
\usepackage{graphicx}
\usepackage{url}
\usepackage{float}
\usepackage{booktabs}
\usepackage{multirow}
\usepackage{graphicx}
\usepackage{bm}
\usepackage{ragged2e}
\usepackage{booktabs}
\usepackage{comment}
\usepackage{array}
\usepackage{ragged2e}
\usepackage{etoolbox}
\usepackage{placeins}
\usepackage{bm}

\usepackage[
  backend=biber,
  style=numeric,          
  citestyle=numeric,      
  sorting=none          
]{biblatex}
\usepackage[colorlinks=true, citecolor=green, linkcolor=blue, urlcolor=blue]{hyperref}

\begin{document}
\title{Brain Tumor Segmentation in the Sub-Saharan African Population Using Segmentation-Aware Data Augmentation and Model Ensembling}
%
%
\author{Claudia Takyi Ankomah\inst{1}\orcidID{0009-0004-7139-6927} \and
Livingstone Eli Ayivor\inst{1} \and
Ireneaus Nyame\inst{2} \and 
Leslie Wambo\inst{1} \and 
Patrick Yeboah Bonsu\inst{3} \and
Aondona Moses Iorumbur \inst{5} \and
Raymond Confidence \inst{4} \and
Toufiq Musah\inst{1}}

 
\institute{Kwame Nkrumah University of Science and Technology, Kumasi, Ghana.\and
University of Cape Coast, Cape Coast, Ghana. \and
University of Toronto, Toronto, Canada \and
Department of Biomedical Engineering, McGill University, Montreal, Canada. \and
Department of Physics, Federal University of Technology, Minna, Nigeria.}
\maketitle
\markboth  {C.T. Ankomah et al.} {Brain Tumor Segmentation in SSA Population}
\begin{abstract}
Brain tumors, particularly gliomas, pose significant chall-enges due to their complex growth patterns, infiltrative nature, and the variability in brain structure across individuals, which makes accurate diagnosis and monitoring difficult. Deep learning models have been developed to accurately delineate these tumors. However, most of these models were trained on relatively homogenous high-resource datasets, limiting their robustness when deployed in underserved regions. In this study, we performed segmentation-aware offline data augmentation on the BraTS-Africa dataset to increase the data sample size and diversity to enhance generalization. We further constructed an ensemble of three distinct architectures, MedNeXt, SegMamba, and Residual-Encoder U-Net, to leverage their complementary strengths. Our best-performing model, MedNeXt, was trained on 1000 epochs and achieved the highest average lesion-wise dice and normalized surface distance scores of 0.86 and 0.81 respectively. However, the ensemble model trained for 500 epochs produced the most balanced segmentation performance across the tumour subregions. This work demonstrates that a combination of advanced augmentation and model ensembling can improve segmentation accuracy and robustness on diverse and underrepresented datasets. Code available at: \url{https://github.com/SPARK-Academy-2025/SPARK-2025/tree/main/SPARK2025_BraTs_MODELS/SPARK_NeuroAshanti}.
\end{abstract}

\section{Introduction}
Brain tumors, particularly gliomas, represent a significant challenge in neuro-oncology due to their infiltrative nature, histological heterogeneity, and variable prognosis \cite{ferreira2024we}. Gliomas originate from glial cells such as astrocytes and oligodendrocytes and are categorized by the World Health Organization (WHO) into grades I to IV based on histological and molecular features \cite{bhimavarapu2024brain}. While low-grade gliomas (LGGs) tend to grow slowly and are sometimes curable via surgery, high-grade gliomas (HGGs), including glioblastoma multiforme (GBM), are aggressive and associated with poor outcomes. Despite standard-of-care therapies, median survival for GBM patients remains approximately 16 months \cite{louis20162016}.\\
\indent These tumors pose additional challenges due to their complex growth patterns and the variability in brain structure across individuals, which makes accurate diagnosis and monitoring difficult. Timely and precise tumor segmentations from magnetic resonance imaging (MRI) scans are important for improving patient outcomes through better treatment planning and response assessment \cite{neetha2024segmentation}. \\
\indent Manual segmentation of gliomas from multimodal MRI is considered the clinical gold standard. It is however time-consuming, labor-intensive, and prone to inter- and intra-observer variability \cite{foster2014review}. To overcome these limitations, deep learning-based approaches have been employed, demonstrating superior performance and efficiency compared to traditional image processing methods \cite{huang2022fully,kamnitsas2017efficient}. The Brain Tumor Segmentation (BraTS) Challenge has played a pivotal role in benchmarking and advancing automated glioma segmentation since 2012, offering a standardized dataset that includes preprocessed T1-weighted, T1Gd, T2-weighted, and T2-FLAIR sequences, with expert annotations for enhancing tumor (ET), necrotic and non-enhancing tumor core (NCR/NET), and peritumoral edema (ED) \cite{menze2014multimodal}. \\
\indent Despite progress, a persistent bottleneck in deploying deep learning models for clinical segmentation remains the limited size and diversity of training datasets. This problem becomes especially pronounced when models are applied to out-of-distribution populations \cite{kelly2019key}. The recently introduced BraTS-Africa dataset begins to address this gap by providing multimodal MRI scans of glioma patients from healthcare institutions in sub-Saharan Africa. It includes 146 annotated cases, 95 adult diffuse gliomas, and 51 other central nervous system neoplasms reviewed by radiologists certified by the American Board of Radiology \cite{torp20222021,adewole2025brats}. The BraTS-Africa dataset is relatively small, and the variability in scanner quality, clinical settings, and imaging protocols presents further challenges. As a result, training highly accurate and robust segmentation models in these settings remains an open problem, especially under data-limited conditions and diverse real-world constraints.

\subsection{Related Works}
Over the past decade, convolutional neural networks (CNNs) have emerged as the predominant approach for brain tumor segmentation, particularly in benchmark challenges such as BraTS \cite{menze2014multimodal,bakas2018identifying}. Among these, U-Net and its many derivatives have become foundational, since the launch of the BraTS challenge in 2014, due to their encoder-decoder architecture with skip connections, which help preserve spatial information critical for accurate medical image segmentation \cite{ronneberger2015u}. To improve performance, these models have been progressively enhanced with architectural innovations such as attention gates, residual connections, and cascaded refinement stages \cite{oktay2018attention,li2022residual}. Over time, top-performing solutions began to integrate ensemble learning techniques by combining predictions from multiple U-Net variants to reduce model variance and enhance generalization \cite{isensee2021nnu, myronenko20183d}. \\
\indent   A major advancement was made with the introduction of nnU-Net. It has been the base model of winning solutions in BraTS 2020 \cite{isensee2020nnu}, 2021 \cite{luu2021extending}, and 2022 \cite{zeineldin2022multimodal}. By 2023, this trend evolved further, with the winning team employing an ensemble of a SwinUNETR and nnU-Net to improve tumor subregion segmentation. Complementary strategies like aggressive augmentation, GAN-generated data, and ensemble-based label refinement have also been used to address class imbalance and reduce overfitting \cite{ferreira2024we}. Modern segmentation models such as MedNeXt \cite{roy2023mednext} have moved beyond the traditional U-net architecture. The model was validated across multiple benchmarks, including BraTS, showing that MedNeXt matches or surpasses nnU-Net in accuracy while maintaining scalability and robustness across diverse tumor types and imaging conditions \cite{roy2023mednext}. Newer approaches such as the State Space based \cite{gu2023mamba} SegMamba have been proposed address the high computational demands of the Transformer-based models like SwinUNETR. \\
\indent Our approach builds on recent advances by exploring both model diversity and advanced data augmentation strategies. We ensemble three distinct architectures; MedNeXt, SegMamba, and Residual-Encoder U-Net (ResEnc U-Net), to leverage the complementary strengths of convolutional neural networks and state-space models. To address limited data sample size and variability, we implement a segmentation-aware offline data augmentation pipeline that combines standard geometric and intensity-based transformations together with a custom label-masked elastic deformation to generate diverse and anatomically plausible samples. The label-masked transform applies aggressive elastic deformations specifically to the tumor regions. This targeted augmentation increases variability within the lesion context, thereby enhancing model generalization in data-limited settings such as those represented in the BraTS-Africa dataset.

\section{Methods}
\subsection{Data Description}
This study utilized the BraTS-Africa dataset from the MICCAI-CAMERA-Lacuna Fund BraTS-Africa 2025 Challenge, which comprises 95 preoperative glioma cases (60 for training and 35 for validation)\cite{adewole2023brain}. Each case consists of four MRI sequences: T1-weighted (T1), T1-Contrast Enhanced (T1-CE), T2-weighted (T2), and T2-Fluid Attenuated Inversion Recovery (T2-FLAIR). These modalities provide complementary structural and pathological details that are important for accurate tumour assessment and segmentation. The corresponding ground truth labels were manually annotated by expert radiologists and define three tumor subregions, namely the Enhancing Tumour (ET), Non-enhancing Tumour Core (TC), and Whole Tumour (WT). ET represents regions with contrast uptake on post-contrast T1 images, TC includes necrotic or cystic components that lack enhancement, and WT corresponds to peritumoral edema, which is visible on FLAIR sequences.

\subsection{Data Expansion}
Given the limited sample size of the dataset, segmentation-aware offline data augmentation pipeline was implemented. This approach was adopted to increase the number of data samples, thereby enhancing model generalizability and robustness across varied anatomical presentations of brain tumours, which are important for brain tumour segmentation.

\begin{figure}
   \centering
    \includegraphics[width=1\linewidth]{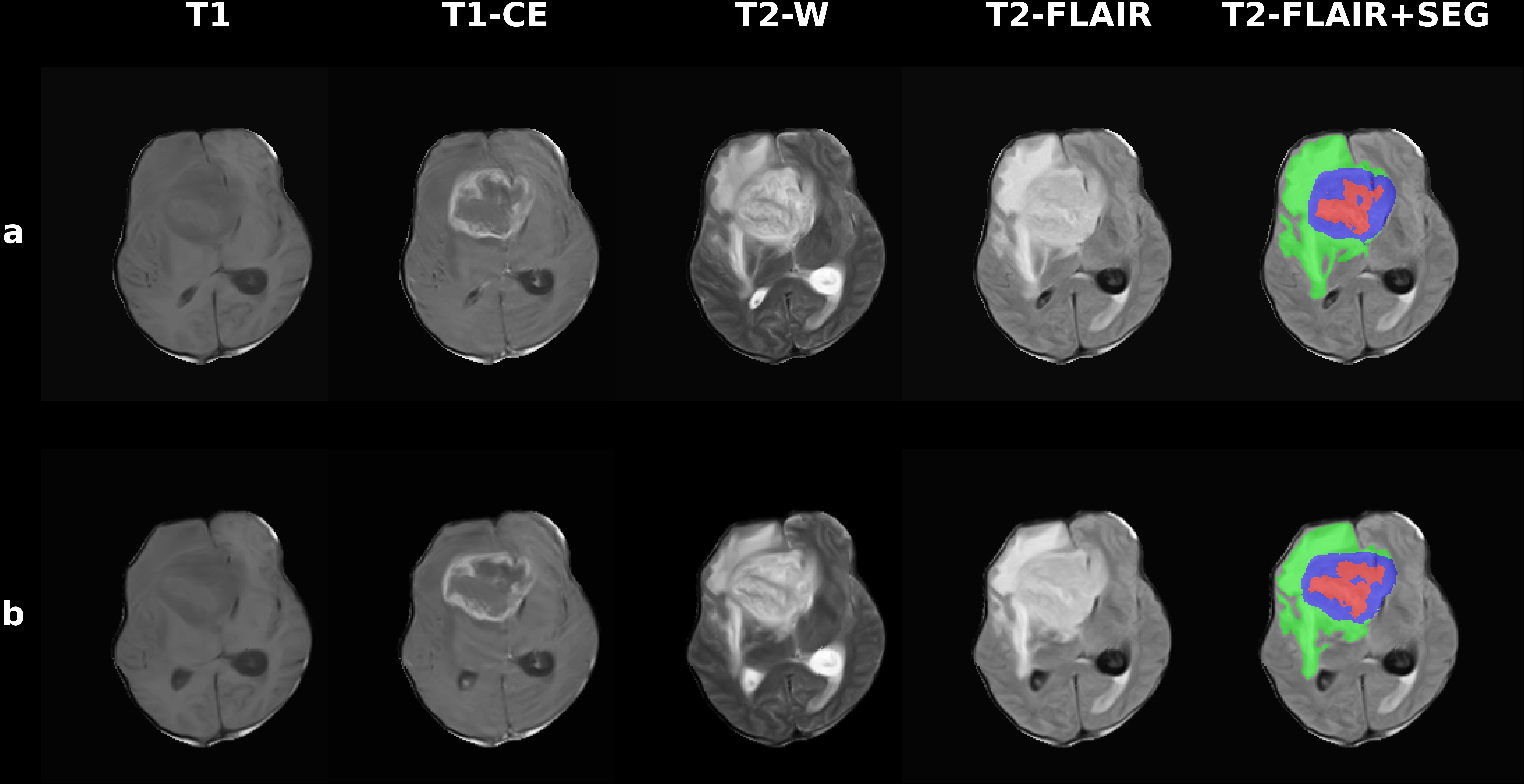}
    \caption{Axial slices of multi-parametric MRI sequences from the BraTS-Africa dataset. (a) Original sample (BraTS-SSA-00007-000).(b) Augmented version of (a) using segmentation-
    aware offline data augmentation. Columns show T1, T1CE, T2-W, T2-FLAIR, and T2-FLAIR with segmentation overlay, where the ET,TC and WT are represented in blue, red, and green respectively.} 
    \label{fig:1}
\end{figure}

\subsubsection{Segmentation-Aware Offline Data Augmentation} 

: This augmentation pipe-line combines affine, flip, bias-field, and elastic deformation transforms from the TorchIO \cite{perez2021torchio} library, and a custom label-mask transform. All the transforms were chained, and predefined probabilities were assigned to all except the label-mask transform, which was applied deterministically. This introduced controlled variation and generation of new samples with realistic brain anatomical characteristics as seen in Figure \ref{fig:1}(b). \\
\indent Complementing the whole brain transformation, the label-mask transform applied rigorous elastic deformation only within the tumour regions of the subject. This preserved the structural integrity of the surrounding brain tissue, enabling targeted morphological changes to the tumour shape and boundary.


 \subsection{Models}
 Multiple segmentation models were selected to explore how variations in their strengths contribute to brain tumour segmentation. All models were trained using the nnU-Net framework to utilize its streamlined pipeline for data preprocessing, training, and inference. Additionally, an ensemble of the selected models was implemented to harness the distinct advantages of each architecture. 
 \subsubsection{SegMamba:}SegMamba is a novel 3D medical image segmentation model that incorporates a U-shaped architecture with Mamba, a state space model designed for long-range dependency modelling \cite{xing2024segmamba}. Unlike transformer-based approaches, SegMamba can efficiently capture both global and multiscale features from an entire input volumetric space with less computational overhead. The architecture consists of a 3D encoder with multiple Tri-orientated Spatial Mamba (TSMamba) blocks for multiscale global feature modelling, a convolution-based decoder, and skip connections with Feature-level Uncertainty Estimation (FUE) modules to refine features from the encoder that are fed to the upsampling path.
 \subsubsection{MedNeXt:}MedNeXt is a fully convolutional segmentation model that employs ConvNeXt blocks in its encoder and decoder, combining the inherent inductive bias of convolutional networks with the scalable design principles of Transformer architectures \cite{roy2023mednext}. The model replaces conventional upsampling and downsampling layers with residual inverted bottlenecks to preserve semantic feature details, leverages an upsampling-based kernel to smoothly transfer pretrained weights from smaller to larger kernels, and supports compound scaling in depth, width, and kernel size for flexible adaptability across diverse segmentation tasks.
 \subsubsection{Residual-Encoder U-NET:}Residual-Encoder U-Net is a derivative of the original nnU-Net framework that introduces residual connections into the encoder pathway \cite{ji2024application}. This architectural network replaces the plain convolutional blocks in the encoder with residual blocks, which reinforce feature representations and gradient flow in deep networks. The decoder structure and overall topology remain consistent with the standard implementation, ensuring compatibility with the self-adapting configuration pipeline of nnU-Net. This modification helps improve learning efficiency and model performance, particularly in deep networks, and it is well-suited for challenging segmentation tasks such as brain tumour segmentation, where detecting subtle variations in tumour appearance is important.

 \subsection{Experiments}
 A series of experiments were conducted to evaluate the performance of our selected models on the BraTS-Africa dataset along with the augmented datasets. Each model was initially trained from scratch for 50 epochs using the original BraTS-Africa dataset. Training continued for 500, and then a 1000 epochs using the best saved model weights checkpoint to assess the impact of extended training durations on segmentation performance. All models were trained in a 5-fold cross validation manner, using a composite loss function of Dice and Cross-Entropy loss to balance region-level and voxel-wise accuracy. Stochastic Gradient Descent (SGD) was employed as the optimizer, with a cosine annealing learning rate schedule initialized at 1e-4. \\
 \indent In creating the offline augmented dataset, transformations were applied 5 times per sample to the training set, forming a total of 300 cases.\\
 \indent Segmentation performance was evaluated using standard metrics, including Lesion-Wise Dice (LSD) and Normalized Surface Distance (NSD) at 1.00 mm for the Enhancing Tumor (ET), Non-enhancing Tumor Core (TC), and Whole Tumor (WT) subregions. Model predictions on the 35 validation cases were submitted to the synapse platform for evaluation.\\

\section{Results}
\subsection{Segmentation Performance}
Tables \ref{tab:results} and \ref{tab:results-ensemble} summarize the segmentation performance of SegMamba, MedNeXt, Residual-Encoder U-Net, and the ensemble across different training configurations. 

\begin{table}[htbp]
\centering
\caption{Results of the experiments comparing individual model performance (LSD - Lesion-Wise Dice and NSD\_1.0 mm - Normalized Surface Distance at 1.00 mm) across the tumor subregions Enhancing Tumor (ET), Tumor Core (TC), and Whole Tumor (WT), as well as their overall average (AVG), under different training durations and settings. Each method is denoted by its initial: \textit{S} for SegMamba, \textit{M} for MedNeXt, and \textit{R} for ResEnc U-Net, followed by the corresponding training configuration. Best results per column are in \textbf{Bold}.}

\begin{tabular}{lcccccccc}
\toprule
\multirow{2}{*}{\textbf{Method}} & \multicolumn{4}{c}{\textbf{LSD}} & \multicolumn{4}{c}{\textbf{NSD 1.0 mm}} \\
\cmidrule(lr){2-5} \cmidrule(lr){6-9}
& ET & TC & WT & AVG & ET & TC & WT & AVG \\
\midrule
\textit{S}\textsubscript{50e} & 0.796 & 0.790 & 0.877 & 0.821 & 0.786 & 0.714 & 0.777 & 0.759 \\
\textit{M}\textsubscript{50e} & 0.799 & 0.794 & 0.879 & 0.824 & 0.793 & 0.729 & 0.790 & 0.771 \\
\textit{R}\textsubscript{50e} & 0.791 & 0.787 & 0.881 & 0.820 & 0.777 & 0.704 & 0.783 & 0.755 \\
\midrule
\textit{S}\textsuperscript{aug}\textsubscript{50e} & 0.799 & 0.785 & 0.871 & 0.818 & 0.761 & 0.770 & 0.699 & 0.743 \\
\textit{M}\textsuperscript{aug}\textsubscript{50e} & 0.804 & 0.790 & 0.887 & 0.827 & 0.782 & 0.715 & 0.776 & 0.758 \\
\textit{R}\textsuperscript{aug}\textsubscript{50e} & 0.791 & 0.783 & 0.857 & 0.810 & 0.835 & 0.770 & \textbf{0.823} & 0.809 \\
\midrule
\textit{S}\textsubscript{500e} & 0.806 & 0.790 & 0.892 & 0.829 & 0.795 & 0.717 & 0.797 & 0.770 \\
\textit{M}\textsubscript{500e} & 0.836 & 0.831 & \textbf{0.912} & 0.860 & 0.789 & 0.725 & 0.778 & 0.764 \\
\textit{R}\textsubscript{500e} & 0.801 & 0.798 & 0.866 & 0.822 & 0.789 & 0.725 & 0.778 & 0.764 \\
\midrule

\textit{S}\textsubscript{1000e} & 0.801 & 0.781 & 0.893 & 0.825 & 0.792 & 0.715 & 0.801 & 0.769 \\
\textit{M}\textsubscript{1000e} & \textbf{0.855} & \textbf{0.844} & 0.896 & \textbf{0.865} & \textbf{0.846} & \textbf{0.775} & 0.808 & \textbf{0.810} \\
\textit{R}\textsubscript{1000e} & 0.800 & 0.796 & 0.864 & 0.820 & 0.790 & 0.726 & 0.778 & 0.765 \\
\midrule
\textit{M}\textsuperscript{aug}\textsubscript{1000e} & 0.836 & 0.838 & 0.876 & 0.850 & 0.831 & 0.775 & 0.775 & 0.794 \\
\bottomrule
\end{tabular}
\label{tab:results}
\end{table}




\begin{table}[htbp]
\centering
\caption{Results of the ensemble experiments (LSD - Lesion-Wise Dice and NSD\_1.0 mm - Normalized Surface Distance at 1.00 mm) across the tumor subregions Enhancing Tumor (ET), Tumor Core (TC), and Whole Tumor (WT), as well as their overall average (AVG). Ensembles are denoted by listing their constituent models: \textit{S} (SegMamba), \textit{M} (MedNeXt), and \textit{R} (ResEnc U-Net). Best results are in \textbf{Bold}.}
 \begin{tabular}{lcccccccc}
 \toprule
 \multirow{2}{*}{\textbf{Ensemble Method}} & \multicolumn{4}{c}{\textbf{LSD}} & \multicolumn{4}{c}{\textbf{NSD 1.0 mm}} \\
 \cmidrule(lr){2-5} \cmidrule(lr){6-9}
 & ET & TC & WT & AVG & ET & TC & WT & AVG \\
 \midrule
\textit{S}+\textit{M}+\textit{R}\textsubscript{50e} & 0.795 & 0.787 & 0.881 & 0.821 & 0.787 & 0.717 & 0.791 & 0.765 \\
\textit{S}+\textit{M}+\textit{R}\textsubscript{500e} & 0.846 & 0.843 & 0.896 & 0.861 & 0.839 & 0.775 & 0.812 & 0.809 \\
\textit{S}+\textit{M}+\textit{R}\textsubscript{1000e} & 0.857 & 0.842 & 0.883 & 0.861 & 0.847 & 0.772 & 0.798 & 0.806 \\
\textit{S}+\textit{M}\textsubscript{1000e} & 0.824 & 0.810 & 0.895 & 0.843 & 0.814 & 0.740 & 0.806 & 0.787 \\
\textit{M}+\textit{R}\textsubscript{1000e} & \textbf{0.860} & \textbf{0.846} & 0.897 & \textbf{0.867} & \textbf{0.852} & \textbf{0.780} & 0.812 & \textbf{0.815} \\ 
\textit{M}+\textit{R}+\textit{M}\textsuperscript{aug}\textsubscript{1000e} & 0.848 & 0.843 & \textbf{0.898} & 0.863 & 0.843 & \textbf{0.780} & \textbf{0.816} & 0.813 \\
 \bottomrule
 \end{tabular}
 \label{tab:results-ensemble}
\end{table}

\begin{figure}[htbp]
    \centering
    \includegraphics[width=1\linewidth]{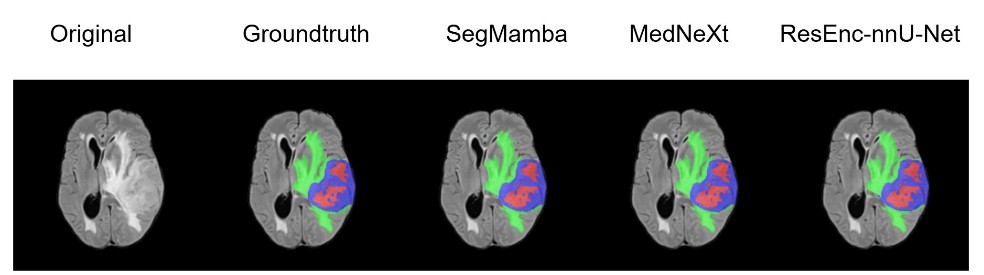}
    \caption{Predictions of the three tumour subregions (TC in red, ET in blue, WT in green) by each of the models. Order of images: Original, Ground truth, SegMamba, MedNeXt, ResEnc U-Net, from sample BraTS-SSA-00141-000.}
    \label{fig:enter-label}
\end{figure}

\section{Discussion and Conclusion}
This study investigated the segmentation performance of three distinct deep learning models, MedNeXt, SegMamba, and ResEnc U-Net and their ensemble, on the BraTS-Africa dataset under varying training setups. The findings reveal that the model architecture, augmentation approach, training configuration and integration strategy play a significant role in determining segmentation accuracy, particularly within data-limited and anatomically diverse cases. \\
\indent Training duration had a notable impact on model performance. Across all architectures, extending training epochs generally led to improved segmentation results. This trend was apparent in the MedNeXt model, which recorded the highest average LSD of \textbf{0.865} and NSD of \textbf{0.810} at 1000 epochs (\textit{M}\textsubscript{1000e}). Interestingly, the best WT segmentation performance in terms of LSD, with a score of  \textbf{0.912}, was achieved by the MedNeXt-model trained for 500 epochs (\textit{M}\textsubscript{500e}). This suggests that optimal performance on individual  tumour subregions may not always correlate linearly with increased training duration. In contrast, SegMamba showed diminishing results beyond 500 epochs, suggesting that optimal training duration may vary depending on architectural characteristics and convergence behavior. \\
 \indent MedNeXt consistently outperformed the other models across both LSD and NSD metrics, likely due to its ability to preserve spatial information through its ConvNeXt residual blocks and efficient scaling. SegMamba demonstrated competitive performance as well, particularly for WT segmentation. This may be due to its state-space modelling and better global context encoding capabilities. The ResEnc U-Net showed stable but slightly lower performance, which may be attributed to its relatively simpler architectural design.\\ 
 \indent The segmentation-aware augmented dataset yielded NSD improvements for some models compared to their baselines. The \textit{R}\textsuperscript{aug}\textsubscript{50e} recorded NSD scores of \textbf{0.835}, \textbf{0.770}, and \textbf{0.823} for ET, TC, and WT respectively, with its WT score being the highest among all models and its average NSD increasing from \textbf{0.755} to \textbf{0.809}. \textit{S}\textsuperscript{aug}\textsubscript{50e} showed a notable increase in TC NSD, although its overall average NSD declined slightly. Meanwhile, \textit{M}\textsuperscript{aug}\textsubscript{50e} exhibited a small decrease in average NSD, suggesting that augmentation may not consistently benefit all architectures. While its effects varied across models, the observed improvements highlight the benefit of targeted augmentation in improving boundary-level segmentation performance.\\
\indent The ensemble models consistently outperformed their individual counterparts in both LSD and NSD metrics, confirming the benefit of combining diverse architectural representations. The \textit{M}+\textit{R}\textsubscript{1000e}  configuration achieved the highest average NSD of \textbf{0.815} and tied for the highest LSD of \textbf{0.867}. The Residual-Encoder U-Net was initially excluded from the ensemble (results indicated by \textit{S}+\textit{M}\textsubscript{1000e}) based on the assumption that its standalone performance was inferior to the other models. However, its inclusion was later shown to enhance overall ensemble performance, suggesting that it contributed meaningfully to the final predictions. Another noteworthy observation was that, although SegMamba achieved competitive standalone performance, its removal from the ensemble unexpectedly led to a slight improvement in overall results. This outcome suggests that not all high-performing models contribute positively when combined, and that ensemble behavior may depend on how different architectures how they interpret MRI modalities differently, as shown by Ren et al.  \cite{ren2025here}. Future work should examine how these differences affect ensemble performance. Collectively, these findings suggest that integrating multiple architectures produces a more balanced and generalizable model and highlights its potential for use in clinical contexts where these attributes are essential.
\\
\indent Although increasing training duration generally improved model performance, the variation in fold usage and training epochs from 50 to 1000 may have led to unstable learning in certain configurations. A more uniform strategy, such as training all models on the same folds and applying a cosine annealing schedule with warm restarts, could have yielded smoother convergence and better generalization. Future work can explore such dynamic scheduling approaches to stabilize training across folds and enhance performance without significantly increasing runtime.\\
\indent Overall, the findings demonstrate that brain tumour segmentation performance is influenced by the type of model architecture, training configuration, augmentation strategy, and ensemble learning. Leveraging these factors collectively in this study supports the development of robust brain tumour segmentation models with improved generalizability, suitable for deployment in Sub-Saharan Africa and other resource-limited medical imaging settings.

\begin{credits}
\subsubsection{\ackname} The authors would like to thank the following instructors of the Sprint AI Training for African Medical Imaging Knowledge Translation (SPARK) Academy 2025 summer school on deep learning in medical imaging for providing insightful background knowledge on brain tumours that informed the research presented here; Noha Magdy, Maruf Adewole, Ayomidale B. Oladele, Amal Saleh, Nourou Dine Bankole, Jeremiah Fadugba, Lorumbur Moses, Toufiq Musah, Teresa Zhu, Craig Jones, Confidence Raymond, Lukman E. Ismaila, Ugumba Kikwima, Mehdi Astaraki, Peter Hastreiter, Evan Calabrese, Esin Uzturk Isik, Navodini Wijethilake, Rancy Chepchirchir, James Gee, MacLean Nasrallah, Jean Baptiste Poline, Bijay Adhikari, Kenneth Agu, Mohannad Barakat \& Yahoo Liu. The authors would also like to thank Linshan Liu for administrative assistance in supporting the SPARK Academy training and capacity-building activities, which the authors immensely benefited from. The authors acknowledge the computational infrastructure support from the Digital Research Alliance of Canada (The Alliance) and the University of Washington Azure GenAI for Science Hub through The eScience Institute and Microsoft (PI: Mehmet Kurt) secured for the SPARK Academy. Finally, we would like to thank the Lacuna Fund for Health and Equity, the Radiological Society of North America (RSNA), the Research \& Education (R\&E) Foundation Derek Harwood-Nash International Education Scholar Grant, the McGill University Healthy Brain and Healthy Lives (HBHL) and the National Science and Engineering Research Council of Canada (NSERC) Discovery Launch Supplement for making the SPARK Academy possible via research grant supports.

\subsubsection{\discintname}
The authors have no competing interests to declare that are relevant to the content of this article
\end{credits}

\printbibliography

@article{ferreira2024we,
  title={How we won brats 2023 adult glioma challenge? just faking it! enhanced synthetic data augmentation and model ensemble for brain tumour segmentation},
  author={Ferreira, Andr{\'e} and Solak, Naida and Li, Jianning and Dammann, Philipp and Kleesiek, Jens and Alves, Victor and Egger, Jan},
  journal={arXiv preprint arXiv:2402.17317},
  year={2024}
}

@article{bhimavarapu2024brain,
  title={Brain tumor detection and categorization with segmentation of improved unsupervised clustering approach and machine learning classifier},
  author={Bhimavarapu, Usharani and Chintalapudi, Nalini and Battineni, Gopi},
  journal={Bioengineering},
  volume={11},
  number={3},
  pages={266},
  year={2024},
  publisher={MDPI}
}

@article{louis20162016,
  title={The 2016 World Health Organization classification of tumors of the central nervous system: a summary},
  author={Louis, David N and Perry, Arie and Reifenberger, Guido and Von Deimling, Andreas and Figarella-Branger, Dominique and Cavenee, Webster K and Ohgaki, Hiroko and Wiestler, Otmar D and Kleihues, Paul and Ellison, David W},
  journal={Acta neuropathologica},
  volume={131},
  number={6},
  pages={803--820},
  year={2016},
  publisher={Springer}
}

@article{neetha2024segmentation,
  title={Segmentation and classification of brain tumour using LRIFCM and LSTM},
  author={Neetha, KS and Narayan, Dayanand Lal},
  journal={Multimedia Tools and Applications},
  volume={83},
  number={31},
  pages={76705--76730},
  year={2024},
  publisher={Springer}
}

@article{foster2014review,
  title={A review on segmentation of positron emission tomography images},
  author={Foster, Brent and Bagci, Ulas and Mansoor, Awais and Xu, Ziyue and Mollura, Daniel J},
  journal={Computers in biology and medicine},
  volume={50},
  pages={76--96},
  year={2014},
  publisher={Elsevier}
}

@article{huang2022fully,
  title={Fully convolutional network for the semantic segmentation of medical images: A survey},
  author={Huang, Sheng-Yao and Hsu, Wen-Lin and Hsu, Ren-Jun and Liu, Dai-Wei},
  journal={Diagnostics},
  volume={12},
  number={11},
  pages={2765},
  year={2022},
  publisher={MDPI}
}

@article{kamnitsas2017efficient,
  title={Efficient multi-scale 3D CNN with fully connected CRF for accurate brain lesion segmentation},
  author={Kamnitsas, Konstantinos and Ledig, Christian and Newcombe, Virginia FJ and Simpson, Joanna P and Kane, Andrew D and Menon, David K and Rueckert, Daniel and Glocker, Ben},
  journal={Medical image analysis},
  volume={36},
  pages={61--78},
  year={2017},
  publisher={Elsevier}
}

@article{menze2014multimodal,
  title={The multimodal brain tumor image segmentation benchmark (BRATS)},
  author={Menze, Bjoern H and Jakab, Andras and Bauer, Stefan and Kalpathy-Cramer, Jayashree and Farahani, Keyvan and Kirby, Justin and Burren, Yuliya and Porz, Nicole and Slotboom, Johannes and Wiest, Roland and others},
  journal={IEEE transactions on medical imaging},
  volume={34},
  number={10},
  pages={1993--2024},
  year={2014},
  publisher={IEEE}
}

@article{kelly2019key,
  title={Key challenges for delivering clinical impact with artificial intelligence},
  author={Kelly, Christopher J and Karthikesalingam, Alan and Suleyman, Mustafa and Corrado, Greg and King, Dominic},
  journal={BMC medicine},
  volume={17},
  number={1},
  pages={195},
  year={2019},
  publisher={Springer}
}

@article{torp20222021,
  title={The WHO 2021 Classification of Central Nervous System tumours: a practical update on what neurosurgeons need to know—a minireview},
  author={Torp, Sverre Helge and Solheim, Ole and Skjulsvik, Anne Jarstein},
  journal={Acta neurochirurgica},
  volume={164},
  number={9},
  pages={2453--2464},
  year={2022},
  publisher={Springer}
}

@article{adewole2025brats,
  title={The BraTS-Africa Dataset: Expanding the Brain Tumor Segmentation Data to Capture African Populations},
  author={Adewole, Maruf and Rudie, Jeffrey D and Gbadamosi, Anu and Zhang, Dong and Raymond, Confidence and Ajigbotoshso, James and Toyobo, Oluyemisi and Aguh, Kenneth and Omidiji, Olubukola and Akinola, Rachel and others},
  journal={Radiology: Artificial Intelligence},
  volume={7},
  number={4},
  pages={e240528},
  year={2025},
  publisher={Radiological Society of North America}
}

@article{bakas2018identifying,
  title={Identifying the best machine learning algorithms for brain tumor segmentation, progression assessment, and overall survival prediction in the BRATS challenge},
  author={Bakas, Spyridon and Reyes, Mauricio and Jakab, Andras and Bauer, Stefan and Rempfler, Markus and Crimi, Alessandro and Shinohara, Russell Takeshi and Berger, Christoph and Ha, Sung Min and Rozycki, Martin and others},
  journal={arXiv preprint arXiv:1811.02629},
  year={2018}
}

@inproceedings{ronneberger2015u,
  title={U-net: Convolutional networks for biomedical image segmentation},
  author={Ronneberger, Olaf and Fischer, Philipp and Brox, Thomas},
  booktitle={International Conference on Medical image computing and computer-assisted intervention},
  pages={234--241},
  year={2015},
  organization={Springer}
}

@article{oktay2018attention,
  title={Attention u-net: Learning where to look for the pancreas},
  author={Oktay, Ozan and Schlemper, Jo and Folgoc, Loic Le and Lee, Matthew and Heinrich, Mattias and Misawa, Kazunari and Mori, Kensaku and McDonagh, Steven and Hammerla, Nils Y and Kainz, Bernhard and others},
  journal={arXiv preprint arXiv:1804.03999},
  year={2018}
}

@article{li2022residual,
  title={Residual-attention UNet++: a nested residual-attention U-net for medical image segmentation},
  author={Li, Zan and Zhang, Hong and Li, Zhengzhen and Ren, Zuyue},
  journal={Applied Sciences},
  volume={12},
  number={14},
  pages={7149},
  year={2022},
  publisher={MDPI}
}

@article{isensee2021nnu,
  title={nnU-Net: a self-configuring method for deep learning-based biomedical image segmentation},
  author={Isensee, Fabian and Jaeger, Paul F and Kohl, Simon AA and Petersen, Jens and Maier-Hein, Klaus H},
  journal={Nature methods},
  volume={18},
  number={2},
  pages={203--211},
  year={2021},
  publisher={Nature Publishing Group}
}

@inproceedings{myronenko20183d,
  title={3D MRI brain tumor segmentation using autoencoder regularization},
  author={Myronenko, Andriy},
  booktitle={International MICCAI brainlesion workshop},
  pages={311--320},
  year={2018},
  organization={Springer}
}

@inproceedings{isensee2020nnu,
  title={nnU-Net for brain tumor segmentation},
  author={Isensee, Fabian and J{\"a}ger, Paul F and Full, Peter M and Vollmuth, Philipp and Maier-Hein, Klaus H},
  booktitle={International MICCAI Brainlesion Workshop},
  pages={118--132},
  year={2020},
  organization={Springer}
}

@inproceedings{luu2021extending,
  title={Extending nn-UNet for brain tumor segmentation},
  author={Luu, Huan Minh and Park, Sung-Hong},
  booktitle={International MICCAI brainlesion workshop},
  pages={173--186},
  year={2021},
  organization={Springer}
}

@inproceedings{zeineldin2022multimodal,
  title={Multimodal CNN networks for brain tumor segmentation in MRI: a BraTS 2022 challenge solution},
  author={Zeineldin, Ramy A and Karar, Mohamed E and Burgert, Oliver and Mathis-Ullrich, Franziska},
  booktitle={International MICCAI Brainlesion Workshop},
  pages={127--137},
  year={2022},
  organization={Springer}
}

@inproceedings{roy2023mednext,
  title={Mednext: transformer-driven scaling of convnets for medical image segmentation},
  author={Roy, Saikat and Koehler, Gregor and Ulrich, Constantin and Baumgartner, Michael and Petersen, Jens and Isensee, Fabian and Jaeger, Paul F and Maier-Hein, Klaus H},
  booktitle={International Conference on Medical Image Computing and Computer-Assisted Intervention},
  pages={405--415},
  year={2023},
  organization={Springer}
}

@article{gu2023mamba,
  title={Mamba: Linear-time sequence modeling with selective state spaces},
  author={Gu, Albert and Dao, Tri},
  journal={arXiv preprint arXiv:2312.00752},
  year={2023}
}

@inproceedings{xing2024segmamba,
  title={Segmamba: Long-range sequential modeling mamba for 3d medical image segmentation},
  author={Xing, Zhaohu and Ye, Tian and Yang, Yijun and Liu, Guang and Zhu, Lei},
  booktitle={International conference on medical image computing and computer-assisted intervention},
  pages={578--588},
  year={2024},
  organization={Springer}
}

@article{adewole2023brain,
  title={The brain tumor segmentation (brats) challenge 2023: Glioma segmentation in sub-saharan africa patient population (brats-africa)},
  author={Adewole, Maruf and Rudie, Jeffrey D and Gbdamosi, Anu and Toyobo, Oluyemisi and Raymond, Confidence and Zhang, Dong and Omidiji, Olubukola and Akinola, Rachel and Suwaid, Mohammad Abba and Emegoakor, Adaobi and others},
  journal={ArXiv},
  pages={arXiv--2305},
  year={2023}
}

@article{karanam2025morph,
  title={MORPH-LER: Log-Euclidean Regularization for Population-Aware Image Registration},
  author={Karanam, Mokshagna Sai Teja and Iyer, Krithika and Joshi, Sarang and Elhabian, Shireen},
  journal={arXiv preprint arXiv:2502.02029},
  year={2025}
}

@inproceedings{kumar2023learning,
  title={Learning Transferable Object-Centric Diffeomorphic Transformations for Data Augmentation in Medical Image Segmentation},
  author={Kumar, Nilesh and Gyawali, Prashnna K and Ghimire, Sandesh and Wang, Linwei},
  booktitle={International Conference on Medical Image Computing and Computer-Assisted Intervention},
  pages={255--265},
  year={2023},
  organization={Springer}
}

@article{perez2021torchio,
  title={TorchIO: a Python library for efficient loading, preprocessing, augmentation and patch-based sampling of medical images in deep learning},
  author={P{\'e}rez-Garc{\'\i}a, Fernando and Sparks, Rachel and Ourselin, S{\'e}bastien},
  journal={Computer methods and programs in biomedicine},
  volume={208},
  pages={106236},
  year={2021},
  publisher={Elsevier}
}

@incollection{ji2024application,
  title={Application of 3D nnU-Net with residual encoder in the 2024 MICCAI head and neck tumor segmentation challenge},
  author={Ji, Kaiyuan and Wu, Zhihan and Han, Jing and Jia, Jun and Zhai, Guangtao and Liu, Jiannan},
  booktitle={Challenge on Head and Neck Tumor Segmentation for MRI-Guided Applications},
  pages={250--258},
  year={2024},
  publisher={Springer}
}

@article{fadugba2025deep,
  title={Deep ensemble approach for enhancing brain tumor segmentation in resource-limited settings},
  author={Fadugba, Jeremiah and Lieberman, Isabel and Ajayi, Olabode and Osman, Mansour and Akinola, Solomon Oluwole and Mustvangwa, Tinashe and Zhang, Dong and Anazondo, Udunna C and Confidence, Raymond},
  journal={arXiv preprint arXiv:2502- .02179},
  year={2025}
}

@article{zhao2024transferring,
  title={Transferring Knowledge from High-Quality to Low-Quality MRI for Adult Glioma Diagnosis},
  author={Zhao, Yanguang and Bai, Long and Zhang, Zhaoxi and Wu, Yanan and Islam, Mobarakol and Ren, Hongliang},
  journal={arXiv preprint arXiv:2410- .18698},
  year={2024}
}

@article{parida2024adult,
  title={Adult glioma segmentation in sub-saharan africa using transfer learning on stratified finetuning data},
  author={Parida, Abhijeet and Capell{\'a}n-Mart{\'\i}n, Daniel and Jiang, Zhifan and Tapp, Austin and Liu, Xinyang and Anwar, Syed Muhammad and Ledesma-Carbayo, Mar{\'\i}a J and Linguraru, Marius George},
  journal={arXiv preprint arXiv:2412.04111},
  year={2024}
}

@article{hashmi2024optimizing,
  title={Optimizing brain tumor segmentation with mednext: Brats 2024 ssa and pediatrics},
  author={Hashmi, Sarim and Lugo, Juan and Elsayed, Abdelrahman and Saggurthi, Dinesh and Elseiagy, Mohammed and Nurkamal, Alikhan and Walia, Jaskaran and Maani, Fadillah Adamsyah and Yaqub, Mohammad},
  journal={arXiv preprint arXiv:2411.15872},
  year={2024}
}

@article{ren2025here,
  title={Here Comes the Explanation: A Shapley Perspective on Multi-contrast Medical Image Segmentation},
  author={Ren, Tianyi and Rivera, Juampablo Heras and Oswal, Hitender and Pan, Yutong and Chopra, Agamdeep and Ruzevick, Jacob and Kurt, Mehmet},
  journal={arXiv preprint arXiv:2504.04645},
  year={2025}
}

\end{document}